\documentclass[%
 reprint,
 superscriptaddress,preprintnumbers,
 nofootinbib,
 amssymb,
 aps,
]{revtex4-1}

\usepackage[utf8]{inputenc}
\usepackage{hyperref}
\usepackage{amsmath,amssymb,mathtools}
\usepackage{dsfont}
\usepackage{graphicx}   
\usepackage[table]{xcolor}
\usepackage{colortbl}
\usepackage{url}
\usepackage[normalem]{ulem}
\usepackage{slashed}
\usepackage[capitalise, english]{cleveref}
\usepackage[noindentafter]{titlesec} 
\allowdisplaybreaks

\usepackage{siunitx}
\usepackage{xspace}

\usepackage{tikz}
\usetikzlibrary{arrows.meta, positioning}

    \makeatletter
    \def\CT@@do@color{%
      \global\let\CT@do@color\relax
            \@tempdima\wd\z@
            \advance\@tempdima\@tempdimb
            \advance\@tempdima\@tempdimc
    \advance\@tempdimb\tabcolsep
    \advance\@tempdimc\tabcolsep
    \advance\@tempdima2\tabcolsep
            \kern-\@tempdimb
            \leaders\vrule
                    \hskip\@tempdima\@plus  1fill
            \kern-\@tempdimc
            \hskip-\wd\z@ \@plus -1fill }
    \makeatother

\usepackage{accents}
\DeclareMathSymbol{\widetildesym}{\mathord}{largesymbols}{"65}

\def\thesubsection{\arabic{section}.\arabic{subsection}}

\def\thesection{\arabic{section}}
\titleformat*{\subsubsection}{\normalfont \small \bfseries \boldmath}
\renewcommand{\paragraph}[1]{\vspace{.3em} \indent {\bfseries \boldmath #1 ---}\xspace }
\makeatletter
    \renewcommand{\p@subsection}{}
    \renewcommand{\p@subsubsection}{}
\makeatother

\definecolor{red}{rgb}{0.6,.0706,.1373}
\definecolor{blue}{rgb}{0,0.396,0.741}
\newcommand\myshade{80}
\colorlet{mylinkcolor}{violet}
\colorlet{mycitecolor}{violet}
\colorlet{myurlcolor}{violet}

\hypersetup{
  linkcolor  = mylinkcolor!\myshade!black,
  citecolor  = mycitecolor!\myshade!black,
  urlcolor   = myurlcolor!\myshade!black,
  colorlinks = true
}
\colorlet{greenref}{green!50!black}

\newcommand{\hc}{\; + \; \mathrm{H.c.} }
\newcommand{\U}{\mathrm{U}}

\newcommand{\LL}{\mathrm{L}}

\newcommand{\dd}{\mathop{}\!\mathrm{d}}

\renewcommand{\Im}{\mathop{\mathrm{Im}}}
\renewcommand{\Re}{\mathop{\mathrm{Re}}}
\newcommand{\eminus}{\vcenter{\hbox{\scalebox{0.6}[1]{$ - $}}}}	

\newcommand{\sscript}[1]{{\scriptscriptstyle \mathrm{#1}}}

\renewcommand{\L}{\mathcal{L}}
\newcommand{\CHDx}{C_{HD}^{\scriptscriptstyle{(\times)}}}


\begin{document}


\title{
  When Two Loops Matter: Electroweak Precision in the SMEFT
}

\author{Lukas Born}
\email{lukas.born@unibe.ch}
\affiliation{Albert Einstein Center for Fundamental Physics, Institute for Theoretical Physics, University of Bern,  Sidlerstrasse 5, CH-3012 Bern, Switzerland}

\author{Admir Greljo}
\email{admir.greljo@unibas.ch}
\affiliation{Department of Physics, University of Basel, Klingelbergstrasse 82,  CH-4056 Basel, Switzerland}

\author{Anders Eller Thomsen}
\email{anders.thomsen@unibe.ch}
\affiliation{Albert Einstein Center for Fundamental Physics, Institute for Theoretical Physics, University of Bern,  Sidlerstrasse 5, CH-3012 Bern, Switzerland}

\date{\today}


\begin{abstract}

We identify a novel next-to-leading order renormalization effect in the dimension-six SMEFT with direct phenomenological impact. The Higgs--Yukawa operator that modifies the top--Higgs coupling $\kappa_t$ induces a shift in the 
$ W $ mass at two-loop order through a large anomalous dimension, rendering electroweak precision observables a powerful indirect probe of $\kappa_t$. We show that this effect is essential for the consistent interpretation of data from future Tera-$Z$ and Giga-$W$ factories such as FCC-ee. The effect is realized in a simple renormalizable two-Higgs doublet model.

\end{abstract}

\maketitle

\section{Introduction} 
\label{sec:intro}

Many extensions of the Standard Model (SM) at high scales first manifest themselves in precision measurements of SM processes. In the precision era, the exceptional experimental and theoretical control over key observables enables sensitivity to even small, loop-induced effects of new physics (NP). Calculations of such effects rely on the use of effective field theories (EFTs), such as the SMEFT~\cite{Buchmuller:1985jz, Giudice:2007fh, Grzadkowski:2010es, Henning:2014wua, Brivio:2017vri, Isidori:2023pyp, Aebischer:2025qhh}, where the loop-induced processes are captured either in the matching step or with operator-mixing during renormalization group (RG) running. Methods and tools for tree-level and one-loop matching~\cite{Aebischer:2023nnv, Dawson:2022ewj, Criado:2017khh, Fuentes-Martin:2020udw, Carmona:2021xtq, Cohen:2020qvb, Fuentes-Martin:2022jrf, Fuentes-Martin:2022vvu, Jenkins:2017jig, Dekens:2019ept, Guedes:2023azv} and one-loop RG running~\cite{Alonso:2013hga, Jenkins:2013wua, Jenkins:2013zja, Jenkins:2017dyc, Machado:2022ozb, Kumar:2021yod} along with one-loop calculations of observables~\cite{Dawson:2019clf, Dawson:2022bxd, Bellafronte:2023amz, Biekotter:2025nln, Degrande:2020evl, Rossia:2024rfo} are now standard ingredients of modern phenomenological frameworks~\cite{Straub:2018kue, Aebischer:2018iyb, DeBlas:2019ehy, Smolkovic:2026cba, EOSAuthors:2021xpv, Allwicher:2022mcg, Giani:2023gfq, Costantini:2024xae} and application studies~\cite{Silvestrini:2018dos, Kley:2021yhn, Greljo:2022jac, Feruglio:2015gka, Davidson:2020hkf, Plakias:2023esq, Delzanno:2024ooj, Calibbi:2017uvl, Panico:2016ull, Aebischer:2020dsw, Brod:2022bww, Grunwald:2025kot, Aebischer:2018csl, deBlas:2025hbr, Armadillo:2026mvp}. 

A recent milestone in the SMEFT program is the completion of the full next-to-leading-order (NLO) two-loop RG running~\cite{Born:2026xkr}, including the baryon number--violating sector~\cite{Banik:2025wpi}. Despite inducing nontrivial correlations among Wilson coefficients absent at leading order, these effects are formally suppressed. This raises a question: is NLO RG evolution phenomenologically relevant for current or future precision experiments?

This question becomes particularly sharp in the context of a future Tera-$Z$ factory. As emphasized in the European Strategy for Particle Physics~\cite{deBlas:2025gyz}, a collider such as FCC-ee~\cite{FCC:2025lpp} would achieve unprecedented precision in the electroweak sector, extending sensitivity to tiny new-physics effects~\cite{Allwicher:2023shc, Allwicher:2024sso, Greljo:2024ytg, Maura:2024zxz, Greljo:2025ggc, Gargalionis:2024jaw, Stefanek:2024kds,  Erdelyi:2024sls, Erdelyi:2025axy, Olgoso:2025jot, Davighi:2024syj, Allwicher:2025mmc}. At this level of accuracy, formally suppressed contributions can become numerically relevant. In this Letter, we identify two-loop SMEFT operator-mixing structures that are crucial to the consistent interpretation of FCC-ee measurements. In particular, \cref{sec:EFT} shows that currently allowed regions of SMEFT parameter space can yield observable effects in electroweak precision observables (EWPOs) at Tera-$Z$ and Giga-$W$ factories via two-loop running. This conclusion is supported by the explicit two-Higgs-doublet model (2HDM) presented in \cref{sec:Model}, where the two-loop contribution will be essential for an interpretation at FCC-ee.

It has previously been observed that EWPOs are sensitive to specific two-loop contributions, notably those involving the mixing of the four-top operator into the Higgs current operator and subsequently into the $T$-parameter operator~\cite{Allwicher:2023aql, Allwicher:2023shc, Stefanek:2024kds, Haisch:2024wnw}. This effect is already captured by the ``square'' of one-loop running~\cite{Jenkins:2013zja, Jenkins:2013wua, Alonso:2013hga}, corresponding to leading-log contributions, and is therefore not specific to genuine two-loop (next-to-leading-log) running. As we demonstrate, however, there exist phenomenologically relevant operator-mixing effects at two-loop order that \emph{cannot} be reproduced within one-loop running.

Only the most sensitive EWPOs can probe two-loop–suppressed new-physics effects at the TeV scale: Suppose that NP generates a tree-level contribution to a Wilson coefficient $C_x$, with $C_x(\Lambda)\sim \Lambda^{-2}$ for $\mathcal{O}(1)$ couplings, where $\Lambda$ denotes the new-physics scale. Even if $C_x$ does not directly contribute to any EWPO, it can generate another coefficient $C_y$ at the $Z$ pole through two-loop running:
\begin{equation}
    C_y(m_Z) \sim - \dfrac{\gamma^{(2)}_{yx}}{(4\pi)^4} C_x(\Lambda) \log \dfrac{\Lambda}{m_Z}\,,
\end{equation}
where $\gamma^{(2)}_{yx}$ is the corresponding entry of the two-loop anomalous dimension matrix. This defines an effective scale for $C_y$ of order
\begin{equation}
    \Lambda_\mathrm{eff} \sim \dfrac{16 \pi^2\, \Lambda}{\sqrt{ |\gamma^{(2)}_{yx}| \log \frac{\Lambda}{m_Z} }}\,. 
\end{equation}
The large loop suppression implies that an EWPO must be sensitive to effective scales of ${\cal O}(100)\,\text{TeV}$ to indirectly probe TeV-scale NP contributing to $C_x$. Nevertheless, the sensitivity to the underlying scale $\Lambda$ is enhanced if $\gamma^{(2)}_{yx}$ is anomalously large; indeed, some entries in the two-loop SMEFT anomalous dimension matrix are $\mathcal{O}(100)$~\cite{Born:2026xkr}. Accounting for this, along with recent analyses~\cite{Maura:2024zxz, Greljo:2024ytg, Celada:2024mcf, Allwicher:2024sso, Greljo:2025ggc}, we conclude that FCC-ee can reach such sensitivity only for a limited set of Wilson coefficients $C_y$ contributing to a few key observables with the highest precision. This motivates our strategy: to identify an operator $C_x$ that mixes into the most tightly constrained directions $C_y$ with sufficiently large anomalous dimensions. We found prominent examples, which we wish to emphasize in this letter.

\section{Two-Loop SMEFT RG for EWPO at Tera-$Z$}
\label{sec:EFT}

\begin{figure}[t]
\centering
\includegraphics[width=.8\columnwidth]{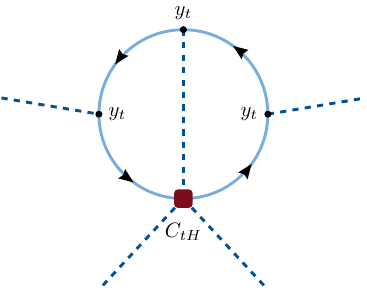}
\caption{Representative two-loop diagram for SMEFT operator mixing, $C_{tH} \to \CHDx$, correlating tree-level modifications of the top Yukawa with $\delta m_W$.}
\label{fig:2loop_CtH_to_CHD}
\end{figure}

We identify modifications to the top Yukawa coupling as a benchmark case for showcasing the significance of NLO RG effects for EWPOs. At leading order, the dimension-six Yukawa operators mix only among themselves, $(C_{uH}, C_{dH}, C_{eH})$, and into $C_H$, with negligible phenomenological impact. At NLO, however, the top-Yukawa operator induces contributions to EWPO-relevant operators enhanced by powers of $y_t$, rendering these effects numerically important, as discussed in \cref{sec:tYuk}.

In \cref{sec:4t}, we revisit the case of the right-handed $4t$ operator contributing to EWPOs at two loops. While the dominant effect is already captured by LO SMEFT running, numerically important corrections arise at NLO.

\subsection{Top-Yukawa and electroweak precision}
\label{sec:tYuk}

Here, we pursue the scenario in a subset of the dimension-six SMEFT to emphasize the effects of interest. The low-scale NP contributions are captured through just three dimension-six operators added to the SM:\footnote{Here and throughout the letter, we employ the `Mainz' basis for the dimension-six SMEFT, which is the one employed for the 2-loop RGEs~\cite{Born:2026xkr}. $ \CHDx $ is the $ C_{HD} $ coefficient of the Warsaw basis~\cite{Grzadkowski:2010es} while $C_{HWB}$ is normalized differently.}
    \begin{equation}
    \begin{split}
    \L_{\sscript{SMEFT}} = &\phantom{+}\CHDx \big(D_\mu H^\dagger H\big) \big(H^\dagger D^\mu H \big) \\
    &+ g_Y g_L \, C_{HWB} (H^\dagger t^I H) W^I_{\mu\nu} B^{\mu\nu} \\
    &+ \Big[ C_{tH} |H|^2\, \overline{q}_3 \widetilde{H} t \hc \Big].
    \end{split}
    \end{equation}
The dimension-six top Yukawa operator (last line) modifies the coupling of the physical Higgs boson to the top quark to 
    \begin{equation}
    y_t^\mathrm{eff} = \frac{\sqrt{2} m_t}{v} + \delta y_t, \qquad  \delta y_t = - v^2 C_{tH}
    \end{equation}
at tree-level, where $ v = \SI{246}{GeV}$ is the Higgs vacuum expectation value (VEV) and $ m_t $ the top quark mass. Any non-zero Wilson coefficient $ C_{tH} $ breaks the SM relation between the top quark mass and its coupling to the Higgs boson. 

Experimental constraints on the top--Higgs interaction are commonly reported in the $\kappa$ framework, defined by
    \begin{equation}
    \kappa_t e^{i\phi_t}=  \dfrac{\delta y_t}{y_t} + 1, \qquad y_t = \frac{\sqrt{2} m_t}{v}.
    \end{equation}
The current experimental status after LHC Run~2 gives $ \kappa_t = \num{0.94(11)} $ (ATLAS~\cite{ATLAS:2022vkf}) and $ \kappa_t =1.01^{+0.11}_{-0.10} $ (CMS~\cite{CMS:2022dwd}) and are in good agreement with the SM prediction $ \kappa_t = 1$; however, sizable NP deviations remains possible. The analysis of Ref.~\cite{Brod:2022bww} implies the bound $|\sin\phi_t\,\kappa_t|\lesssim \num{1.0e-3}$ from the induced contribution to the electron electric dipole moment (EDM). 

The projected precision for HL-LHC is $ |\delta \kappa_t| < 0.034 $~\cite{ATLAS:2025eii}. While FCC-ee is not expected to improve the direct bound on the top-Higgs Yukawa~\cite{deBlas:2025gyz}, we argue that a measurable NP contribution at the HL-LHC generically entails measurable deviations in EWPOs at FCC-ee.

In particular, the $ \CHDx $ and $ C_{HWB} $ coefficients are strongly probed in EWPOs. One might imagine a scenario in which high-scale NP does not directly contribute to these coefficients, effectively turning them off at the NP matching scale. However, the dimension-six top Yukawa operator mixes with the other two operators at NLO, generating an irreducible contribution absent tuned cancellations from other sources. As a suitable benchmark, we consider the scenario where $ \CHDx $ and $ C_{HWB} $ are generated exclusively from the RG evolution of the initial $ C_{tH} $ coupling. The leading contributions to the $\beta$-functions are two-loop order and are given by~\cite{Born:2026xkr}
    \begin{align}
    \dfrac{\dd C_{HD}^{\scriptscriptstyle (\times)}}{\dd \ln \mu} &\supset - \frac{18}{(4\pi)^4}  |y_t|^2 \Re\!\big[ y_t^\ast  C_{tH}\big], \\
    \dfrac{\dd C_{HWB}}{\dd \ln \mu} &\supset - \frac{1}{(4\pi)^4} \Re\!\big[ y_t^\ast  C_{tH}\big].
    \end{align}
The large numerical prefactor in the running of $ \CHDx $ compared to $ C_{HWB} $ results in the former contribution being an order of magnitude more important to the EWPOs.  

The operators $ \CHDx $ and $ C_{HWB} $ contribute to the $ W $ mass in the `LEP' input scheme. The BSM contribution to the $ W $ mass is given by~\cite{Brivio:2020onw}\footnote{with $ s_w \equiv \sin \theta_w $, being the sine of the weak-mixing angle.}
    \begin{equation}
    \begin{split}
    \delta m_W &= -m_W \frac{v^2}{1 -2s_w^2} \left(
    \frac{c_w^2}{4} \CHDx + \frac{e^2}{2} C_{HWB} \right) \\
    &= \SI{16.1}{MeV} \cdot (\cos\phi_t\, \kappa_t- 1) \ln \!\dfrac{\Lambda}{m_Z}\,.
    \end{split}
    \end{equation}
The state-of-the-art CMS determination of the $ W $-mass gives $ \delta m_W = \SI{7(10)}{MeV}$~\cite{CMS:2024lrd} relative to the electroweak fit value. 
Thus, the $W$-boson mass already constrains the top--Higgs coupling at the $\mathcal{O}(1)$ level. These are not yet competitive with direct bounds, given the absence of large logarithmic enhancements in viable scenarios.
By contrast, the two orders of magnitude increase in the experimental sensitivity at FCC-ee, see \cref{tab:EWPO_small}, will constrain $ \delta \kappa_t $ at the percent level.

\begin{table}[t]
    \centering
    \small
    \setlength{\tabcolsep}{4pt}
    \begin{tabular}{lccc}
    \hline\hline
    Observable & S1: TH1+EXP & S2: TH2+EXP & S3: EXP \\
    \hline
    $m_W\, [\si{MeV}]$ & 0.39 & 0.26 & 0.24 \\
    $\sin^2\! \theta_w\, [10^{-6}]$   & 16 
    & 1.8 & 1.7 
    \\
    $A_e\, [10^{-5}]$   & 13 
    & 1.4 
    & 1.3 
    \\
    $R_b\, [10^{-6}]$              & 43  & 3.5  & 0.39 \\
    \hline\hline
    \end{tabular}
    \caption{Projected uncertainties for selected electroweak observables at FCC-ee. Scenarios S1 and S2 include different assumptions on theoretical uncertainties combined with experimental ones taken from~\cite{Greljo:2025ggc, EWPPGTH}, while S3 corresponds to experimental-only uncertainties~\cite{FCC:2025lpp}. }
    \label{tab:EWPO_small}
\end{table}

The operators $ \CHDx $ and $ C_{HWB} $ also contribute to the couplings of the $ Z $-boson and shift the weak mixing angle. Their contribution is given by
    \begin{equation}
    \begin{split}
    \delta s_w^2 &= \frac{v^2}{2 -4s_w^2} \left(
    c_w^2 s_w^2 \CHDx + e^2 C_{HWB} \right) \\
    &= -\num{9.4e-5} \cdot (\cos\phi_t\, \kappa_t- 1) \ln \!\dfrac{\Lambda}{m_Z}.
    \end{split}
    \end{equation}
The measurement obtained by combining forward--backward and left--right asymmetries in the LEP and SLC experiments across leptons and heavy quarks~\cite{ALEPH:2005ab} gives $ \delta s_w^2 = \num{-20(160)e-6} $ relative to the SM best fit prediction~\cite{Reina:2025suh}. FCC-ee is expected to improve precision by (one) two orders of magnitude, assuming (conservative) optimistic theoretical progress (see \cref{tab:EWPO_small}), thereby making $s_w^2$ a powerful complementary probe to $m_W$.

\subsection{Four-top and electroweak precision}
\label{sec:4t}
Another example highlighting the importance of two-loop effects in EWPOs is provided by the four-right-handed-top operator,
    \begin{equation}
    \L_{\sscript{SMEFT}} = C_{tt} \big(\overline{t} \gamma_\mu t\big)^{\!2}.
    \end{equation}
Owing to the absence of top quarks in the proton parton distribution functions, this operator is among the most challenging four-fermion interactions to probe directly at the LHC~\cite{DiNoi:2025uhu, CMS:2025rug}. The CMS interpretation of $t\bar t t\bar t$ production sets the $2\sigma$ bound $|C_{tt}| < 1.2\,\text{TeV}^{-2}$~\cite{CMS:2025rug}. Nevertheless, it is well known that $C_{tt}$ mixes at one loop into the Higgs--top current, which subsequently feeds into $\CHDx$, thereby inducing a leading-logarithmic ($\ln^2$), two-loop contribution to $m_W$~\cite{Allwicher:2023aql, Allwicher:2023shc, Stefanek:2024kds, Haisch:2024wnw}. Remarkably, already at current precision, this indirect effect provides a competitive constraint on $C_{tt}$.

\begin{figure}[t]
    \centering
    \includegraphics[width=\columnwidth]{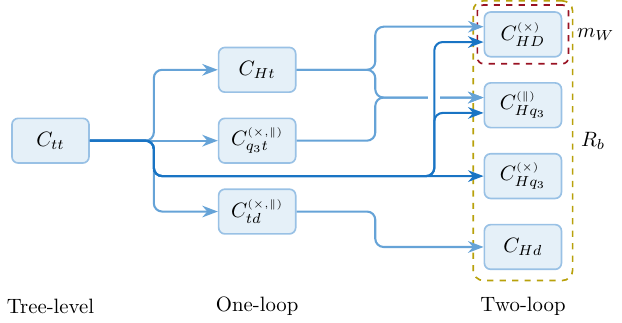}
    \caption{Dimension-six Wilson coefficients induced by the four right-handed top operator, organized by the perturbative order at which they are generated.  Arrows indicate the RG flow contributing to WCs at LO, while dashed boxes highlight the operators entering $m_W$ and $R_b$, respectively. The contributions to operators with right-handed down quarks ($d$) are gauge-mediated and universal across all families.}
    \label{fig:Ctt_flow}
\end{figure}

Another powerful probe of new physics is the ratio
    \begin{equation} \label{eq:Rb_def}
    R_b = \frac{\Gamma(Z \to b \bar{b})}{\Gamma(Z \to \mathrm{hadrons})},
    \end{equation}
which compares the partial width into $b$ quarks to the total hadronic $Z$ width. The current LEP measurement yields $ \delta R_b = \num{410(660)e-6}$ relative to the electroweak fit~\cite{Reina:2025suh,ALEPH:2005ab}. Looking ahead, FCC-ee is expected to improve the sensitivity to $m_W$ by a factor of $\sim 40$, whereas the uncertainty on $R_b$ is projected to shrink by a much larger factor of $\sim 190$ in scenario S2~\cref{tab:EWPO_small}. Notably, even this projection is dominated by theoretical uncertainties, leaving significant room for further improvement. In the experimental-only S3 scenario, the gain of an additional order-of-magnitude may be achievable.

The total contribution of $ C_{tt} $ to $ R_b $ is determined by a mixture of two-loop running into $  C_{Hq_3}^{\scriptscriptstyle(\times,\|)} $ and $ \CHDx $, and several one-loop squared contributions, as illustrated in~\cref{fig:Ctt_flow}. As previously, we integrate the flow down to the EW scale, keeping all two-loop contributions.\footnote{We have also kept terms proportional to SM gauge couplings, giving universal contributions across all quark families (e.g.,  in the down--Higgs current $C_{Hd}$).} Working at next-to-leading logarithm (NLL), we also include the NLO determination of the observables using the results of~\cite{Biekotter:2025nln}, which are provided purely numerically. The relevant NLO contributions to $ R_b $ are exactly the operators that appear in the intermediate step in the `one-loop squared' RG running. All told, we find   
    \begin{equation}
    \delta R_b = \SI{e-6}{TeV^2} \cdot \left(\! 13 \ln^2 \! \dfrac{\Lambda}{m_Z} - 21 \ln \dfrac{\Lambda}{m_Z} \right) C_{tt}\,.
    \end{equation}
For $C_{tt} = \Lambda^{-2}$, FCC-ee probes scales up to $\Lambda \sim \SI{4}{TeV}$ through $R_b$ in scenario S2, with potential to reach $\SI{18}{TeV}$ in the experimental-only scenario S3. Interestingly, the subleading logarithm induces a sizable correction to the leading result (40\% at \SI{5}{TeV}), although unfortunately serving to make the overall correction smaller. 

Thus, $R_b$ provides a complementary handle to $m_W$, which remains the most sensitive probe of $C_{tt}$. For comparison, we present the full two-loop NLL result for $m_W$, obtained within the same framework used for $R_b$:
    \begin{equation}
    \delta m_W = \SI{1}{MeV\cdot TeV^2} \cdot \left(\!  16 \ln \dfrac{\Lambda}{m_Z} - 10 \ln^2 \dfrac{\Lambda}{m_Z} \right) C_{tt}\,.
    \end{equation}
Under the projected FCC-ee uncertainties in scenario S2, sensitivity to $C_{tt}$ probes scales up to $ \SI{20}{TeV}$. Notably, a large cancellation between the leading and subleading log (when running from the TeV scale), will weaken the present bound obtained from only including leading-log running.

\section{Top-philic 2HDM}
\label{sec:Model}

To demonstrate the effects of the NLO running in a realistic BSM model for top-Yukawa modifications, we consider a top-philic 2HDM. A heavy copy of the SM Higgs field ${\Phi\sim(1,2)_{\frac{1}{2}}}$, with mass $M_\Phi$, couples exclusively to the top quark and the SM Higgs doublet:
\begin{align}
\begin{split}
    \L = \L_\mathrm{SM} + \L_{\mathrm{kin}}^\Phi &- \Big[ \lambda_\Phi (\Phi^\dagger H) (H^\dagger H)\\
    &+ y_\Phi  \Bar{q}_3 \widetilde{\Phi} t_R \hc \Big] + \ldots
\end{split}\label{eq:PhiLag}
\end{align}
By neglecting additional Yukawa couplings, in particular those to down-type quarks, and assuming a sufficiently heavy doublet, our setup captures the main features of the decoupling limit of the type-II 2HDM at low $\tan\beta$~\cite{Branco:2011iw}, where the phenomenology is dominated by top-quark interactions.

\paragraph{Matching and running down to the EW scale}
Integrating out $\Phi$ at tree-level generates three SMEFT Wilson coefficients in the Mainz basis~\cite{Born:2026xkr}:
    \begin{equation} \label{eq:tree-matching}
    \begin{split}
    C_{tH}(M_\Phi) &= \frac{1}{M_\Phi^2}\lambda_\Phi^\ast y_\Phi \, , \\
    C_{q_3t}^{\scriptscriptstyle (\times)}(M_\Phi) &= -\dfrac{|y_\Phi|^2}{2M_\Phi^2}, \\
    C_H(M_\Phi) &= \dfrac{|\lambda_\Phi|^2}{M_\Phi^2} .
    \end{split}
    \end{equation}
The dimension-six top-Yukawa coupling is supplemented by contributions to the Higgs self-coupling and a four-top interaction. Due to the constraints on the imaginary part of $ C_{tH} $, we will restrict our analysis to real $ \lambda_\Phi, y_\Phi $ for simplicity.  

At NLO, the model matches onto many more Wilson coefficients. The operators most relevant for the leading EWPOs are Higgs-current operators, electroweak top-quark dipoles, and the kinetic Higgs operator:
\begin{align}
    C_{Ht}(\mu) &= \dfrac{1}{(4\pi)^2} \dfrac{|y_t|^2 |y_\Phi|^2}{M_\Phi^2} \!\left( \ln \!\dfrac{M_\Phi}{\mu} - \dfrac{7}{12} \right), \label{eq:2HDM_Ht} \\
    C_{Hq_3}^{\scriptscriptstyle(\|)}(\mu) &= \dfrac{1}{(4\pi)^2} \dfrac{|y_t|^2 |y_\Phi|^2}{M_\Phi^2} \!\left(\dfrac{2}{3}- \ln \!\dfrac{M_\Phi}{\mu}  \right), \label{eq:cHu1_matching} \\
    C_{Hq_3}^{\scriptscriptstyle(\times)}(\mu) &= -\dfrac{1}{(4\pi)^2} \dfrac{1}{6} \dfrac{|y_t|^2 |y_\Phi|^2}{M_\Phi^2}, \label{eq:cHu2_matching} \\
    C_{tB} (\mu) &= - \dfrac{1}{(4\pi)^2} \dfrac{9}{16} \dfrac{|y_\Phi|^2 y_t}{M_\Phi^2}\,, \\
    C_{tW} (\mu) &= - \dfrac{1}{(4\pi)^2} \dfrac{19}{24} \dfrac{|y_\Phi|^2 y_t}{M_\Phi^2}\,, \\
    C_{HD}^{\scriptscriptstyle(\|)}(\mu) &= -\dfrac{3}{(4\pi)^2} \dfrac{ |\lambda_\Phi|^2}{M_\Phi^2}\,, \label{eq:2HDM_HD}
    \end{align}
at the renormalization scale $\mu$.
The expressions~(\ref{eq:2HDM_Ht}--\ref{eq:2HDM_HD}) include one-loop matching (determined with the \texttt{Matchete} package~\cite{Fuentes-Martin:2022jrf}) and contributions from the linear log running of the tree-level-generated operators (or, equivalently, matching at the scale $ \mu $ rather than $ M_\Phi $).

\begin{figure}[t]
    \centering
    \includegraphics[width=\columnwidth]{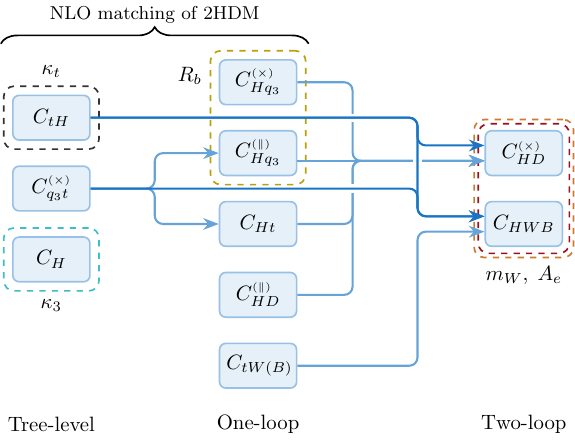}
    \caption{Dimension-six Wilson coefficients induced by the top-philic 2HDM, organized by the perturbative order at which they are generated. Arrows indicate the RG flow contributing to WCs at LO, while dashed boxes highlight the operators contributing to each observable at LO.}
    \label{fig:2HDM_operators}
\end{figure}

The $\CHDx$ and $C_{HWB}$ operators, identified in the previous section, receive contributions from the two-loop running of $ C_{tH} $ and $ C_{q_3t}^{\scriptscriptstyle(\times)} $~\cite{Born:2026xkr},  and the one-loop running from all the one-loop generated operators~(\ref{eq:2HDM_Ht}--\ref{eq:2HDM_HD}) (as wells as a numerically small one-loop matching contribution to $\CHDx$). How the different operators contribute is illustrated in Fig.~\ref{fig:2HDM_operators}. Running down to the weak scale, keeping only the leading contribution from each combination of UV couplings, we obtain
    \begin{align}
    \CHDx(m_Z) =\,& \dfrac{-1}{(4\pi)^2} \dfrac{g_Y^4}{120 M_\Phi^2} + \dfrac{1}{(4\pi)^4} \dfrac{1}{M_\Phi^2} \bigg[ 10g_Y^2 |\lambda_\Phi|^2 \nonumber \\
    &+|y_t|^4 |y_\Phi|^2\! \left(\! 24 \ln\! \dfrac{M_\Phi}{m_Z} -28 \! \right)   \nonumber \\
    &+ 18 |y_t|^2 \Re\! \big(y_t^\ast \lambda_\Phi^\ast y_\Phi\big) \!  
    \bigg] \!\ln \!\dfrac{M_\Phi}{m_Z} \, , \label{eq:2HDM_CHDx}\\
    C_{HWB}(m_Z) =\,& \dfrac{1}{(4\pi)^4} \dfrac{\tfrac{1}{3}|y_t|^2 |y_\Phi|^2+ \Re\! \big(y_t^\ast \lambda_\Phi^\ast y_\Phi\big)}{M_\Phi^2} \ln\! \dfrac{M_\Phi}{m_Z}\, . \label{eq:2HDM_CHWB}
    \end{align}
The terms proportional to $ \Re\! \big(y_t^\ast \lambda_\Phi^\ast y_\Phi\big) $ are due to two-loop running. Additionally, there is a cancellation at the level of $\tfrac{44}{45}$ in the first term in~\eqref{eq:2HDM_CHWB} between the two-loop running from $ C_{q_3t}^{\scriptscriptstyle (\times)} $ and one-loop running from the one-loop matched EW dipoles. 

In addition to the quark and bosonic operators described here, the model induces various leptonic operators at two-loop order, which contribute to $ m_W $ in input schemes such as the LEP~\cite{Biekotter:2025nln}. These operators are suppressed by gauge couplings and by modest numerical coefficients relative to those in~\eqref{eq:2HDM_CHDx} and~\eqref{eq:2HDM_CHWB}, and have only a minor impact on our analyses. Nevertheless, we have included the operators listed in App.~\ref{app:leptonic_operators}.

While matching and running capture important NP effects in the EWPOs, consistently working to NLO also requires determining the observables at one-loop order. Generally speaking, if a WC gives a one-loop running effect to a WC that enters an observable at LO then that operator will itself contribute to the observable at one-loop order. \cref{fig:2HDM_operators} lists all the WCs that are the origins of one-loop flow (light blue arrows). We use the numerical expressions from~\cite{Biekotter:2025nln} at the renormalization scale $ \mu = m_Z $ to capture these effects (see also~\cite{Dawson:2019clf}). Thus, our analysis captures all effects at the leading loop order for each observable, but we do not perform proper RG resummation. This is not expected to qualitatively affect our conclusions.

\paragraph{Additional indirect constraints}
In addition to giving contributions to the $ W $ mass and the weak angle, as anticipated in \cref{sec:tYuk}, the model also induces corrections to the $Z b \bar{b}$, probed by the $ R_b $ ratio~\eqref{eq:Rb_def}. 
This was identified as the leading indirect probe of the 2HDM sector of MSSM at FCC-ee~\cite{Greljo:2025ggc}. The one-loop generated Higgs currents $ C_{Hq_3}^{\sscript{(\|),\times}} $ contribute at LO through~\cite{Biekotter:2025nln}
\begin{equation} \label{eq:RbMainz}
    \delta R_{b} = 24 \frac{\left(9-18s_{w}^2+20s_{w}^4\right) \! \left(3- 2s_w^2\right) }{\left(45-84s_{w}^2+88s_{w}^4\right)^2} v^2 \! \left[C_{Hq_3}^{\sscript{(\|)}} + C_{Hq_3}^{\sscript{(\times)}} \right] .
\end{equation}
As emphasized in \cref{sec:4t}, the ultimate sensitivity of this probe at FCC-ee will depend strongly on theory progress.

The trilinear Higgs-boson self-coupling $\kappa_3$ is a strong probe of the quartic coupling $ \lambda_\Phi $. Its shift is dominated by the tree-level matching contribution to the dimension-six Higgs self-coupling $C_H$~\eqref{eq:tree-matching} through
\begin{align}
    \delta \kappa_3 = -\frac{2 v^4}{m_h^2} C_H\,.
\end{align}
ATLAS obtained a 95\% confidence interval of $\delta \kappa_3 \in [-2.2, 6.2]$~\cite{ATLAS:2024ish} (with a slightly broader interval reported by CMS~\cite{CMS:2024awa}). A future combination of HL-LHC and FCC-ee measurements are expected to bring the sensitivity to $ \delta \kappa_3 $ down to $ 18\% $~\cite{FCC:2025lpp}. For a general discussion of $\delta \kappa_3$ in new physics contexts, see~\cite{Durieux:2022hbu}.

\begin{figure*}
\includegraphics[width=.8\textwidth]{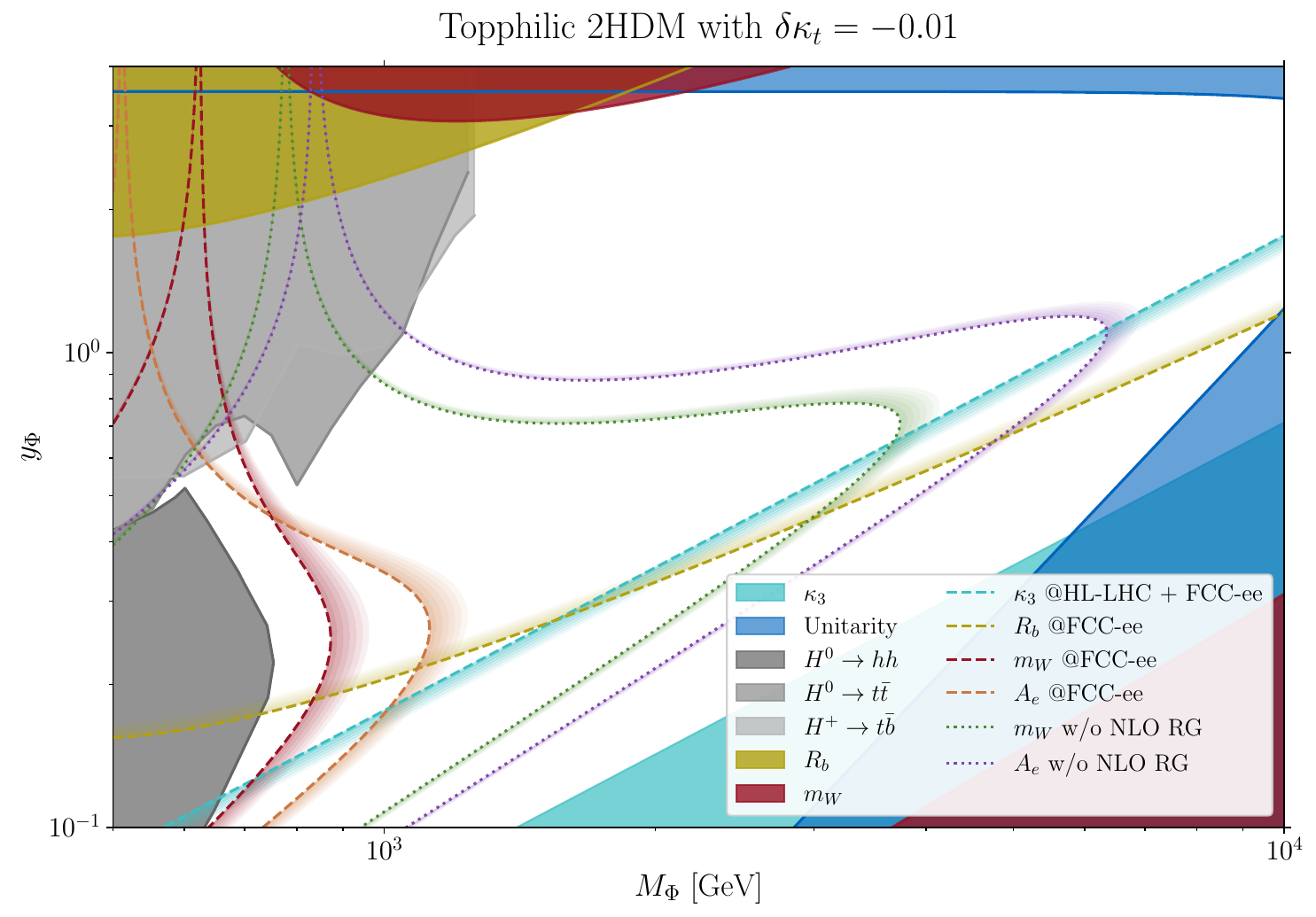}\\[1em]
\includegraphics[width=.8\textwidth]{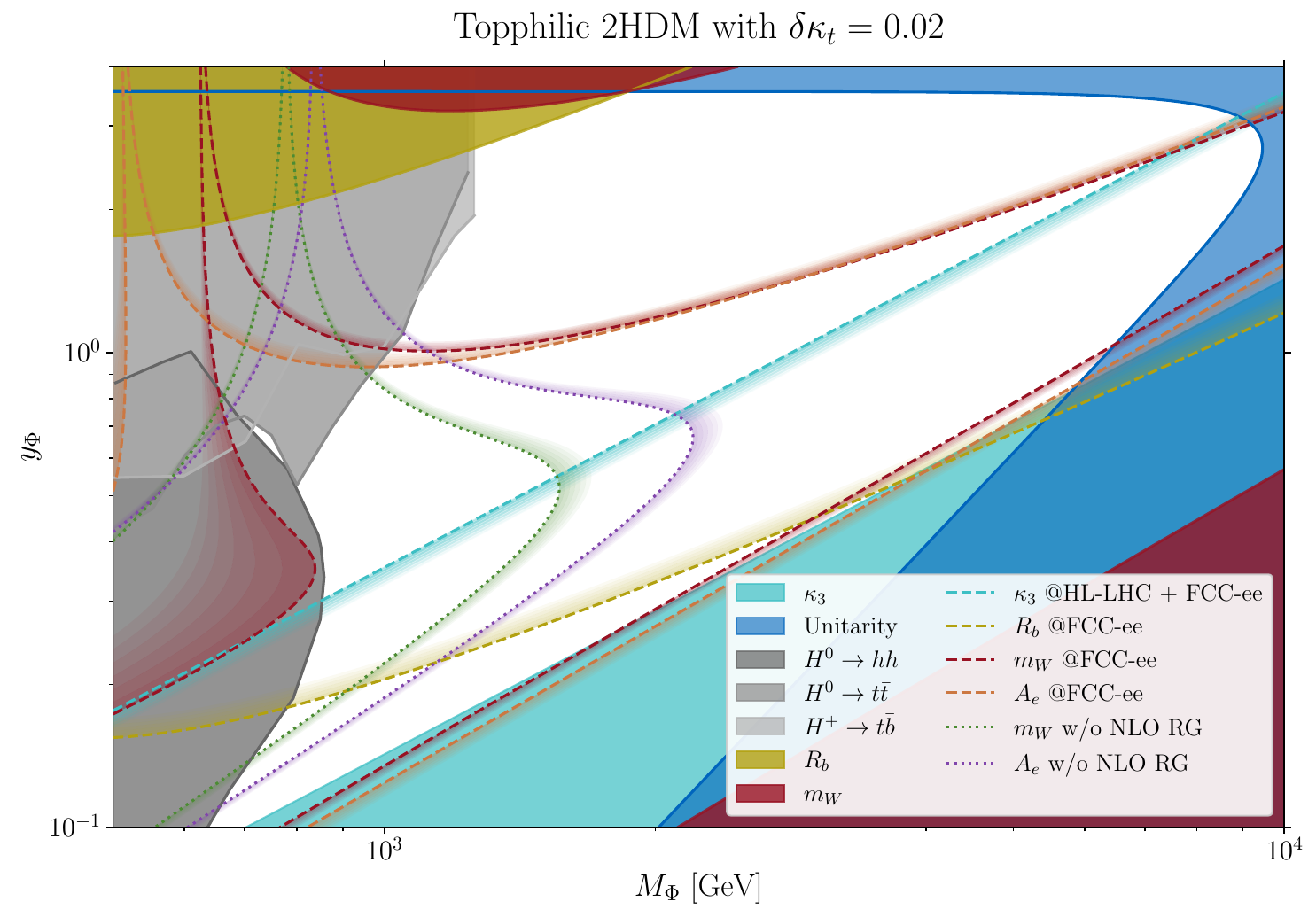}
    \caption{Parameter space for the top-philic 2HDM consistent with $ \delta \kappa_t = -0.01 $ (top) and $ \delta \kappa_t = 0.02 $ (bottom). The filled regions are excluded by the current experimental bounds. The dashed lines indicate the projected FCC-ee sensitivity, whereas the dotted lines indicate the sensitivity in the absence of NLO running. A fading gradient has been drawn on the side of projections that can be probed to help guide the eyes.}\label{fig:MainPlot}
\end{figure*}

\paragraph{Collider searches}  
In the decoupling limit, $ M_\Phi \gg v$, the heavy Higgs doublet decomposes into a nearly degenerate spectrum consisting of a charged scalar $H^{\pm}$ and two neutral states, the CP-even $H^0$ and CP-odd $A$, with masses $m_{H^\pm}\simeq m_{H^0} \simeq m_A \simeq M_\Phi$.\footnote{Absorbing a mixed mass term to write  
$V \supset \lambda_{\Phi} (\Phi^\dagger H)(H^\dagger H - \tfrac{v^2}{2}) + \mathrm{H.c.}$
ensures that $\Phi$ does not acquire a VEV at tree level.} The CP-even $ H^0 $ will mix with the SM Higgs-boson $ h $ with a mixing angle 
    \begin{equation}
    \theta \simeq \lambda_\Phi v^2 / M_\Phi^2 = - \frac{y_t}{y_\Phi} \delta \kappa_t\,,
    \end{equation}
which is parametrically small in the region of interest to our analysis (cf. Fig.~\ref{fig:MainPlot}). This, in particular, ensures the validity of the SMEFT framework~\cite{Dawson:2023ebe, Banta:2023prj, Kraml:2025fpv}.  

The collider phenomenology of the new heavy scalars is controlled by the common mass $ M_\Phi$ and the two couplings in~\eqref{eq:PhiLag}. Single production of the neutral states is dominated by gluon fusion, mediated by the top-quark Yukawa interaction $y_\Phi$. The decay pattern is dictated by the interplay between $y_\Phi$ and $\lambda_\Phi$: for sizable $y_\Phi$, both $H^0$ and $A$ decay predominantly to $t\bar t$, while for sufficiently large (and real) $\lambda_\Phi$, the CP-even state $H^0$ can have a non-negligible branching ratio into di-Higgs ($H^0\to hh$). In the heavy $M_\Phi$ limit, we find 
    \begin{equation}
    \dfrac{\Gamma(H^0\to hh)}{\Gamma(H^0 \to t\bar t\,)} \simeq \dfrac{3 \lambda_\Phi^2 v^2}{2 y^2_\Phi M^2_\Phi}\,,
    \end{equation} 
suggesting the dominance of $H^0 \to t \bar t$ channel for small mixing angles $\theta \ll 1$. 
Since $M_\Phi \gg m_W$, the Goldstone equivalence theorem implies that decays into longitudinal gauge bosons are controlled by the quartic interactions, yielding
    \begin{equation}
    \Gamma(H^0\to WW) : \Gamma(H^0\to ZZ) : \Gamma(H^0\to hh) = 2 : 1 : 9 \,.
    \end{equation}
Similarly, the charged Higgs decay $H^\pm \to t b$ dominates over $H^\pm \to W^\pm h$ for a small $\theta$.

Important direct model constraints arise from searches for heavy resonances in $t\bar t$ final states. In particular, the ATLAS collaboration has performed a search using $140\,\mathrm{fb}^{-1}$ of data at $\sqrt{s}=13\,\mathrm{TeV}$~\cite{ATLAS:2024vxm}. We reinterpret their Fig.~13\,(a), obtained in the type-II 2HDM in the decoupling limit (closely matching our setup), by rescaling the branching fractions to account for the different coupling structure. The resulting $95\%$ CL exclusion provides the dominant direct constraint in our parameter space in the regime of large $y_\Phi$ and small $ M_\Phi$. Another complementary probe at the LHC is the single charged-Higgs production in association with top and bottom quarks, $pp \to H^+ \bar t b + \text{c.c.}$, followed by the decay $H^\pm \to t b$. The corresponding ATLAS search~\cite{ATLAS:2021upq} yields constraints that are numerically comparable to those from the neutral resonances $H^0, A \to t \bar t$ discussed above, as illustrated in Fig.~7\,(top left panel). We reinterpret this result in our setup and report the resulting $95\%$ CL exclusion.

Additional probes include charged Higgs pair production, $pp \to H^+H^-$~\cite{Ahmed:2025jjl}, and resonant di-Higgs production via $H^0 \to hh$~\cite{ATLAS:2023vdy}. The latter is particularly relevant at intermediate values of $y_\Phi$, where the branching ratio into $hh$ is sizable while gluon-fusion production remains efficient.
We reinterpret the $95\%$ CL upper limits on the production cross section from Fig.\,(1b) of the ATLAS search~\cite{ATLAS:2023vdy}. For the signal prediction, we use the heavy Higgs gluon-fusion cross sections from Ref.~\cite{LHCHiggsCrossSectionWorkingGroup:2011wcg}, rescaled by $|y_\Phi/y_t|^2$ and multiplied by the corresponding branching ratios discussed above.
Combining all direct searches, we find that the region $M_\Phi \lesssim 0.6\,\text{TeV}$ is robustly probed across the full range of $y_\Phi$ of interest.

\paragraph{Parameter space}
Our central result is illustrated in~\cref{fig:MainPlot}, which shows benchmark regions of the parameter space that yield $\delta\kappa_t = -0.01$ and $\delta\kappa_t = 0.02$, which are just below the projected HL-LHC sensitivity. These choices fix $ \lambda_\Phi $ as a function of $ M_\Phi $ and $ y_\Phi $ (all parameters assumed real). The plots include present bounds and future sensitivity for the observables discussed so far. We focus on the parameter space $ M_\Phi \geq \SI{500}{GeV} $ and $ y_\Phi \geq 0.1 $, where we have a reasonable separation of scales and small Higgs mixing, as required for employing SMEFT, and which is also constrained by direct searches as discussed above. 

The upper ranges of $ M_\Phi $ and $ y_\Phi $ are governed by perturbative unitarity (blue bound), which we crudely approximate with $ y_\Phi^2, \lambda_\Phi < 4\pi $.\footnote{We take some artistic freedom and actually plot the bound $ y_\Phi^4 + \lambda_\Phi^2 < 16\pi^2 $ in \cref{fig:MainPlot} as a simple interpolation.} Another useful indication of the validity of perturbative calculation comes from the total width-to-mass ratio, which determines whether the heavy scalar remains a reasonably narrow resonance. In the limit $M_\Phi \gg m_t$, the dominant contributions give
\begin{equation}
\frac{\Gamma_{H^0}}{M_\Phi}\simeq \frac{3 y_\Phi^2}{16 \pi} + \frac{3 \lambda_\Phi^2 v^2}{8 \pi M_\Phi^2}\, .
\end{equation}
Requiring, for instance, $\Gamma_{H^0} / M_\Phi \lesssim 0.5$ leads to a slightly stronger constraint on $y_\Phi$, while the bound on $\lambda_\Phi$ remains comparatively weaker.

The gray bounds in the left part of the plot correspond to the heavy-Higgs collider search bounds, which do not constrain masses beyond $ \SI{1250}{GeV}\, (\SI{650}{GeV})$ for $y_\Phi = 2 \,(0.1)$. Interestingly, even though the experimentally allowed interval for the Higgs self-coupling ($\kappa_3$) remains large, it already provides the dominant bound (filled teal region) for intermediate range $ M_\Phi $ and small $ y_\Phi $ (corresponding to large $ \lambda_\Phi $). Improvements in sensitivity at HL-LHC and FCC-ee are expected to cover a significant portion of the currently allowed parameter space. The present bounds on $ R_b$ lead to a largely negligible constraint (filled yellow region). By contrast, it can probe the majority of the parameter space at FCC-ee  (delineated by the dashed yellow line) with aggressive assumptions for theory progress (Scenario 2 in Tab.~\ref{tab:EWPO_small}), as anticipated in~\cite {Greljo:2025ggc}.

\paragraph{NLO RG relevance} Our main focus is on the observables $ m_W $ and $ s_w $, which we identified as being affected by $ C_{tH} $ through two-loop running in \cref {sec:EFT}. Since $ s_w $ is not itself an observable, we use the left-right electron asymmetry $ A_e $ as a good proxy such that we can employ the NLO calculation of the observable~\cite{Biekotter:2025nln}. We observe in \cref{fig:MainPlot} that the current constraint for $ m_W $\footnote{Current $ s_w$ bounds are even weaker.} (filled red) are not dominant in any part of the parameter space. By contrast, the future FCC-ee sensitivity (delineated with dashed red/orange lines) is expected to probe most of the $ \delta \kappa_t=-0.01$ parameter space with the exception of the lower-left corner. On the other hand, a region along the diagonal will not be probed for positive $ \delta \kappa_t$ as demonstrated in the $ \delta \kappa_t= 0.02$ plot. This is primarily due to a cancellation between the terms in the second and third lines of~\eqref{eq:2HDM_CHDx}. This feature disappears entirely if we turn off the two-loop running (the region delineated by the green/purple dotted lines). The opposite behavior is observed in the $ \delta \kappa_t=-0.01$ case where the region of sensitivity shrinks significantly if we turn off the two-loop running. We conclude that ignoring the effects of the two-loop running would lead to a \emph{complete misinterpretation of the data} over a large fraction of the parameter space.

\paragraph{Flavor physics} 
Flavor observables are parametrically protected by the minimally broken $\U(2)_q \times \U(2)_u$ flavor symmetry~\cite{Faroughy:2020ina, Greljo:2022cah}, under which the dominant interaction involves only the third generation: $y_\Phi \bar q_3 \widetilde{\Phi} t_R$. Irreducible flavor violation arises from breaking in the left-handed sector, encoded by the CKM elements forming a doublet spurion $V_q \equiv (V_{td}, V_{ts})^T$, corresponding to a misalignment between the interaction and mass bases. 

The leading probe is $b\to s\gamma$, induced at one loop by the charged Higgs and top quark. In our setup, this contribution matches onto the same dipole operator structure as the SM $W$--top loop, and thus amounts to a universal correction to the Wilson coefficients $C_{7,8}$, parametrically suppressed as $\Delta C_{7,8}/C_{7,8}^{\rm SM}\sim y_\Phi^2\,m_W^2/M_\Phi^2$. The precise bound depends on additional assumptions about the flavor structure and vanishes in the down-alignment limit, where the charged scalar does not couple to $s_\LL$. The flavor alignment can be controlled without tuning in a framework such as minimal flavor protection~\cite{Greljo:2025mwj, Biondini:2026ryb}.

For completeness, in the 
up-aligned limit, we obtain EW dipoles from NLO matching to the SMEFT~\cite{Fuentes-Martin:2022jrf}, which, at low energies, contribute to $C_{7}$. From the bounds on $C_{7}$ from~\cite{Paul:2016urs} we obtain $M_\Phi / |y_\Phi| \gtrsim \SI{500}{GeV} $, which is essentially the region already excluded by direct searches and current EWPO, cf. Fig.~\ref{fig:MainPlot}. This is consistent with expectations from well-studied type-II 2HDMs, where stronger constraints arise due to the additional couplings to down-type quarks~\cite{Czaja:2024tcv, Misiak:2017bgg}.

\section{QCD $\theta$-angle and top Yukawa phase}
Unrelated to EWPOs, we identify a curious effect arising from NLO SMEFT RG, including a novel contribution to the topological QCD $\theta$-angle. This effect involves two SMEFT operators 
    \begin{multline}
        \L_{\sscript{SMEFT}} \supset g_s^2 C_{H\widetilde{G}} |H|^2 G_{\mu\nu} \widetilde{G}^{\mu\nu} \\
        + \left( g_s C_{tG} \overline{q}_3 \sigma^{\mu\nu} t^A \widetilde{H} t G^A_{\mu\nu} \hc \right). 
    \end{multline}
After electroweak symmetry breaking, the CP-odd operator $ C_{H\widetilde{G}} $ gives an additive contribution to the $\theta$-angle operator
    \begin{equation}
        \L \supset \frac{\theta_G}{32\pi^2} G_{\mu\nu} \widetilde{G}^{\mu\nu} \,. 
    \end{equation} 
We identify an RG flow from $C_{tH}$ into $C_{H\widetilde{G}}$ via $C_{tG}$ at NLO~\cite{Born:2026xkr}:
    \begin{align}
        \dfrac{\dd C_{tG}}{\dd \ln \mu} &\supset \frac{3}{(16\pi^2)^2} |y_t|^2 C_{tH} \, , \\
        \dfrac{\dd C_{H\widetilde{G}}}{\dd \ln\mu} &\supset -\frac{4}{16\pi^2} \Im\!\big[y_t^\ast C_{tG} \big] \, .
    \end{align}
This effect generates a non-zero $C_{H\widetilde{G}}$ at the top mass scale, sensitive to the phase of the Higgs coupling:
    \begin{align}
        C_{H\widetilde{G}}(m_t) \supset -\frac{6}{(16\pi^2)^3} |y_t|^2 \Im\!\big[y_t^\ast C_{tH} \big] \ln^2\frac{M_\Phi}{m_t} \, .
    \end{align}
We therefore obtain a correction to the $\theta$-angle of the form
    \begin{equation}\label{eq:theta_angle_final}
    \delta \theta_G = 5.08\cdot10^{-4} \cdot \sin\phi_t\, \kappa_t \,\ln^2\! \frac{M_\Phi}{m_t} \, .
    \end{equation}
Even a small misalignment between the dimension-four and -six top-Yukawa couplings will give a sizable contribution.

The experimental limit on the neutron EDM~\cite{Abel:2020pzs} corresponds to a limit on the effective angle $|\overline{\theta}| \equiv \theta_G -\arg \det (Y_u Y_d)\lesssim 10^{\eminus 10} $ (see e.g.~\cite{Bonanno:2025wcv,Benabou:2025viy}). This is seven orders of magnitude less than the prefactor in~\eqref{eq:theta_angle_final}. Naively, the resulting bound on $ \sin\phi_t\, \kappa_t $ is four orders of magnitude better than the one obtained from the electron EDM~\cite{Brod:2022bww}. One should keep in mind that the bound established through the $ \theta $-angle is model-dependent; the presence of a QCD axion can naturally set $ \overline{\theta} = 0 $ regardless of other NP contributions~\cite{Peccei:1977hh,GrillidiCortona:2015jxo,DiLuzio:2020wdo}. In other words, barring unnatural cancellations, a top-Yukawa phase in the interval $ 10^{\eminus 7} \lesssim \kappa_t \sin\phi_t < 10^{\eminus 3}$ requires the presence of an axion.

\section{Conclusion}
\label{sec:conclusion}

We have identified a genuine two-loop RG effect in the dimension-6 SMEFT, whereby the top-Yukawa operator $C_{tH}$ mixes into the electroweak precision operators $\CHDx$ and $C_{HWB}$. This contribution is absent in one-loop evolution and cannot be reproduced by iterated leading-log effects. Despite the formal loop suppression, the large anomalous dimensions and the exceptional projected precision of future electroweak measurements render this effect phenomenologically relevant. In particular, it induces observable shifts in $m_W$ and $\sin^2\theta_w$ at FCC-ee, even for values of $\delta \kappa_t$ that are beyond the direct reach of future-projected HL-LHC data, establishing EWPOs as sensitive indirect probes of top-Higgs interactions in the Tera-$Z$/Giga-$W$ era (\cref{sec:tYuk}). Additionally, NLO RG effects are required for accurate predictions of $4t$ contributions to EWPOs (\cref{sec:4t}). 

We have further demonstrated that the correlation between modified top--Yukawa coupling and electroweak precision operators arises in a simple renormalizable ultraviolet completion, namely a top-philic two-Higgs-doublet model, over viable regions of parameter space consistent with present constraints (\cref{sec:Model}). In this setting, the genuine two-loop contribution plays a decisive role in the accurate interpretation of future data: depending on the sign of $\delta\kappa_t$, it can either enhance or partially compensate other contributions, but in all cases it induces sizable shifts in the predicted sensitivity. Including this effect is \emph{essential} for the correct interpretation of FCC-ee data across much of the viable parameter space. Our results call for the systematic inclusion of next-to-leading-order SMEFT RG effects~\cite{Born:2026xkr} in future global SMEFT analyses and highlight the increasing importance of theoretical precision in the SM prediction for key observables such as $m_W$, $\sin^2\theta_w$, and $R_b$.

\section*{Acknowledgments}

We thank Ben Stefanek for useful discussions. This work has received funding from the Swiss National Science Foundation (SNF) through the Ambizione grant ``Matching and Running: Improved Precision in the Hunt for New Physics,'' project number 209042, and the program “Swiss
High Energy Physics for the FCC” (CHEF) supported in part by SERI.

\section*{End note}

During the finalization of this manuscript, Ref.~\cite{Mantani:2026fao} appeared and studies the impact of the NLO SMEFT RG~\cite{Born:2026xkr} in global fits.

\appendix 
\renewcommand{\thesection}{\Alph{section}}
\renewcommand{\thesubsection}{\Alph{section}.\arabic{subsection}}
\setcounter{section}{0}

\section{Leptonic operators in the top-philic 2HDM}
\label{app:leptonic_operators}
Several leptonic operators are generated at one- and two-loop order in the top-philic 2HDM discussed in \cref{sec:Model}. They enter the observables $m_W$ and $R_b$. 

From one-loop matching and running, we obtain the following expressions for semi-leptonic coefficients:
\begin{align}
    C^{\sscript{(\|)}} _{\ell_i q_3}(\mu) &= \dfrac{1}{(4\pi)^2} \dfrac{|y_\Phi|^2}{M_\Phi^2} \! \left( \dfrac{35 g_Y^2 }{216} - \dfrac{g_L^2}{72} - \dfrac{2g_Y^2}{9} \ln\! \dfrac{M_\Phi}{\mu} \right), \\
    C^{\sscript{(\times)}} _{\ell_i q_3}(\mu) &= \dfrac{1}{(4\pi)^2} \dfrac{1}{36} \dfrac{g_L^2 |y_\Phi|^2}{M_\Phi^2} , \\
    C_{\ell_i t}(\mu) &= \dfrac{1}{(4\pi)^2} \dfrac{g_Y^2 |y_\Phi|^2}{M_\Phi^2} \! \left( \dfrac{5 }{108}- \dfrac{1}{9} \ln\! \dfrac{M_\Phi}{\mu} \right), \\ 
    C_{q_3 e_i}(\mu) &= \dfrac{1}{(4\pi)^2} \dfrac{g_Y^2 |y_\Phi|^2}{M_\Phi^2} \! \left( \dfrac{35 }{108}- \dfrac{4}{9} \ln\! \dfrac{M_\Phi}{\mu} \right) \, , \\
    C_{e_i t}(\mu) &= \dfrac{1}{(4\pi)^2} \dfrac{g_Y^2 |y_\Phi|^2}{M_\Phi^2} \! \left( \dfrac{5 }{54}- \dfrac{2}{9} \ln \!\dfrac{M_\Phi}{\mu} \right) \, .
\end{align}
Being generated through gauge couplings, these operators are universal across all lepton species. These one-loop generated operators then feed into leptonic Higgs currents and $C_{\ell\ell}$ through through one-loop running, generating a two-loop effect:
\begin{align}
    C^{\sscript{(\times)}}_{H \ell_i}(\mu) =\,& \dfrac{1}{(4\pi)^4} \dfrac{g_L^2}{2M_\Phi^2} \left[|\lambda_\Phi|^2 + |y_t|^2 |y_\Phi|^2 \right] \ln \!\dfrac{M_\Phi}{\mu}, \\
    C^{\sscript{(\|)}} _{H \ell_i}(\mu) =\,& \dfrac{1}{(4\pi)^4} \dfrac{1}{M_\Phi^2} \bigg[- \dfrac{g_L^2 + g_Y^2}{4} |\lambda_\Phi|^2 \nonumber \\
    &\qquad - \big(\tfrac{1}{4} g_L^2 + \tfrac{13}{12}g_Y^2 \big) |y_t|^2 |y_\Phi|^2 \nonumber \\
    &\qquad+ \dfrac{2g_Y^2}{3} |y_t|^2 |y_\Phi|^2 \ln\! \dfrac{M_\Phi}{\mu} \bigg] \ln\! \dfrac{M_\Phi}{\mu}, \\
    C_{He_i}(\mu) =\,&  \dfrac{-1}{(4\pi)^4} \dfrac{g_Y^2}{2M_\Phi^2} \bigg[|\lambda_\Phi|^2 + \tfrac{13}{3} |y_t|^2 |y_\Phi|^2 \nonumber \\
    &\qquad-\frac{8}{3}|y_t|^2 |y_\Phi|^2 \ln \!\dfrac{M_\Phi}{\mu} \bigg] \ln \!\dfrac{M_\Phi}{\mu}, \\
    C_{\ell\ell}^{1221}(\mu) =\,&  \dfrac{-1}{(4\pi)^4} \dfrac{g_L^4}{M_\Phi^2} \dfrac{|y_\Phi|^2}{18} \ln \!\dfrac{M_\Phi}{\mu}.
\end{align}
We have neglected numerically small effects, such as terms proportional to the four powers of the electroweak gauge couplings, arising from NLO matching.

The $ C^{\sscript{(\|, \times)}}_{H \ell_i} $ and $ C_{\ell\ell}^{1221} $ coefficients contribute to the Fermi constant and propagate from there to the prediction of $ m_W $ and $ s_w $ in the LEP input scheme. Additionally, the leptonic Higgs currents contribute directly to the $Z$ coupling to electrons, thereby modifying predictions for $A_e$.

\bibliographystyle{JHEP}
\bibliography{refs.bib}

\end{document}